\documentclass{article}[12pt]

\usepackage{amsfonts,graphicx}

\usepackage[ruled]{algorithm2e}
\renewcommand{\algorithmcfname}{ALGORITHM}
\SetAlFnt{\small}
\SetAlCapFnt{\small}
\SetAlCapNameFnt{\small}
\SetAlCapHSkip{0pt}
\IncMargin{-\parindent}

\begin{document}


\centerline{\LARGE\textbf{Programming Plantation Lines on Driverless Tractors}}
\ \\

\centerline{\large Antonio Elias Fabris$^1$}
\smallskip
\centerline{\large Marcelo Zanchetta do Nascimento$^2$}
\smallskip
\centerline{\large Val\'erio Ramos Batista$^3$}
\bigskip
\begin{flushleft}
{$^1$\footnotesize\it IME-USP, r. Mat\~ao 1010, 05508-090 S\~ao Paulo-SP, Brazil; \tt aef@ime.usp.br}

{\tt https://sites.google.com/site/aefabris}

{$^2$\footnotesize\it FACOM-UFU, av. Jo\~ao Naves de \'Avila 2121, Bl.A, 38400-902 Uberl\^andia-MG, Brazil; \tt nascimento@facom.ufu.br}

{\tt http://www.facom.ufu.br/\~{}nascimento}

{$^3$\footnotesize\it CMCC-UFABC, r. Sta. Ad\'elia 166, Bl.B, 09210-170 St. Andr\'e-SP, Brazil; \tt vramos1970@gmail.com}

{\tt https://sites.google.com/site/vramos1970}
\end{flushleft}

\begin{abstract}
Recent advances in Agricultural Engineering include image processing, robotics and geographic information systems (GIS). Some tasks are still accomplished manually, like drawing plantation lines that optimize productivity. Herewith we present an algorithm to find the optimal plantation lines in linear time. The algorithm is based upon classical results of Geometry which enabled a source code with only 573 lines. We have implemented it in Matlab for sugar cane, and it can be easily adapted to other crops like coffee, maize and soy.  
\end{abstract}







\section{Introduction}
\label{intro}

The spread of technological resources have been improving almost all human activities: Education, Medicine, Architecture, etc. Regarding the sectors of the Economy, they are all increasingly dependent on cutting-edge technology. From top to bottom, they run from the Quinary Sector (top executives in government, healthcare, media) to the Primary Sector (mining, farming, fishing, agriculture).

Any economic activity is classified in one of these sectors. For instance, Engineering belongs to the Secondary Sector. By the way, we would like to comment on Electronic Engineering for two reasons: it is the source of Technology itself, and also an example in which Technology is self-applied. For instance, electric circuits designed through Very-large-scale integration (VLSI) are composed of {\it millions} of terminals. It is impracticable to design and test these circuit projects without the help of specific software. See \cite{Davis} for an example.

In the case of Agriculture there are innumerous examples in which modern technology is applied. For instance, robotics \cite{Tanigaki}, image processing \cite{Bakker,Giudiceandrea} and GIS \cite{Maynard,Picoli}. 

However, some tasks are still accomplished manually. This is the case of finding the ideal plantation lines for certain crops like sugar cane, maize, coffee and soy. Sugar cane is particularly important for two reasons. First because its annual production is concentrated by three leading countries: Brazil, India and China. In million of tonnes they yield ca. 700, 300 and 100, respectively. Second because this crop is an important source of ethanol fuel as a substitute to petrol.

\subsection{Ideal Plantation Lines}
\label{ipl}

Herewith we present an algorithm to draw ideal plantation lines for sugar cane. It can be adapted to other crops, but its implementation was specially devoted to programming the plantation lines on driverless tractors that groove, plant and harvest the sugar cane. In order to optimize productivity, the following conditions must be satisfied:
\begin{enumerate}
\item the tractor cannot slant up or down more than $5^\circ$ while driving;
\item the tractor cannot make curves that are under $50m$ in radius;
\item two neighbouring grooves must be $3m$ distant from each other;
\item there must be the least possible number of plantation lines;
\item on a sloping land the extremities of each plantation line must lie above some of its interior stretches.
\end{enumerate}

Such conditions are based on machine and agricultural restrictions. Conditions (1) and (2) prevent the tractor from turning over, whereas (3) and (4) yield the maximum amount of sugar cane within the plot that minimize tractor man\oe uvres from one plantation line to the other. These man\oe uvres are performed at the boundary of the plot. Finally, condition (5) guarantees that rain water will flow to the inside of the plot. Namely, on a sloping land plantation lines should {\it never} coincide with level curves. Otherwise the water will puddle in the grooves. 

\subsection{Main Contributions of this Paper}
\label{mcp}

To the best of our knowledge, an algorithm that solves the proposed problem was never published before. One of the reasons come from conditions (2) and (3). Since plantation lines are parallel, we can begin with some of them that satisfy (3) but further parallels risk having osculating circles with radii lesser than $50m$. So back-and-forth tests are required and they normally result in a tedious manual process of trial-and-error. We reduce the problem to a robust algorithm which has linear time complexity.

Another reason is that in practice about $50\%$ of the plots are {\it concave} regions. On the one hand, we have obtained a programme that works automatically for convex plots, and the users intervene only for concave plots, in which case they have to decide very little. This characterizes a semi-supervised algorithm. On the other hand, we believe that an unsupervised algorithm, if it ever can be found, would unnecessarily increase the complexity of the programme. The difficulty in developing such a method would make the ratio cost/benefit considerably high.

Finally, although condition (3) implies that the manual drawing of a single plantation line is enough to determine all of them, the optimization problem cannot be solved automatically unless we equate the terrain. Its topography is given as a datafile of coordinates, which are graphically rendered by commercial software of Computer-Aided Design (CAD). Our algorithm includes a quick 2D-interpolation of these coordinates and we work on a terrain equated by the graph of a polynomial $z(x,y)$. By the way, it is worthwhile to mention that most CAD/CAM commercial softwares devoted to agriculture can easily integrate our programme.

\section{Preparing the Input Data}
\label{prep}

\subsection{Obtaining the Topography}
\label{tpg}

Typically, GIS can be used to locate points in a region. However, in the case of plantations we need more accurate data. Hence GIS is useful for fixing a standard meridian and parallel of the Earth, and they will be referential for the $x$ and $y$ coordinates, respectively.

Precise $(x,y,z)$ coordinates of the terrain are then obtained by an unmanned aerial vehicle (UAV). They are stored in a datafile named {\tt level\_curves<n>.txt}, where {\tt n} labels a specific terrain. Our programme is compressed in the file {\tt tml.zip} that also contains $5$ different examples of terrains. Some consecutive lines of {\tt level\_curves3.txt} are shown in Table~\ref{tab:one}.

\begin{table}%
\centerline{Some Consecutive Lines of a Terrain Datafile\label{tab:one}}{%
\ \\
\begin{tabular}{lcl}
510662.36 &	7920042.53 &	100.00 \\
510658.26 & 	7920032.13 &	100.00 \\
510700.36 & 	7920086.57 &	99.00  \\
510689.00 &	7920058.88 &	99.00  \\
510681.14 &	7920035.64 &	99.00  \\
510668.97 &	7920004.84 &	99.00  \\
510664.13 &	7919995.51 &	99.00  \\
510662.18 &	7919992.77 &	99.00  \\
510722.89 &	7920088.81 &	98.00  \\
510707.74 &	7920051.87 &	98.00  \\
\end{tabular}}
\end{table}%

From Table~\ref{tab:one} we see that level curves are ordered by the representing height.

The plot boundary is in fact a polygon, of which the $(x,y)$-vertices are stored in {\tt plot<n>.txt} ({\tt n} must refer to the same previous number). Our file {\tt tml.zip} also contains the corresponding $5$ different plots. A plot-file is like a level curves-file, except for having far fewer points and only two columns (the representing height is omitted).

\subsection{Equating the Terrain}
\label{eqterr}

There are many well-known algorithms devoted to 2D-polynomial interpolation. In Matlab the function {\tt interp2} makes use of some methods as the $C^2$ cubic spline (see \cite{DeBoor}) and the $C^1$ cubic convolution (see \cite{Keys}). They result in smooth 2D-functions that are piecewise defined by 2D-polynomials.

Usually the input level curves-files obtained by the UAV contain points in hundreds for an accurate numerical description of the terrain. However, we just need a sample of these points because a plantation area has to be quite regular. Indeed, in practice any plot can be confined to a square of edgelength $700m$. By calling the highest and the lowest level of a plot $H$ and $L$ respectively, then we must have $0\le H-L\le 40m$. 

Moreover, the plot ought not to be hilly and therefore both ways along the plot boundary from $H$ to $L$ should be monotonically decreasing. This last condition is however just recommendable. We shall see some plots that violate it.

\begin{figure}
\centerline{\includegraphics[scale=0.5]{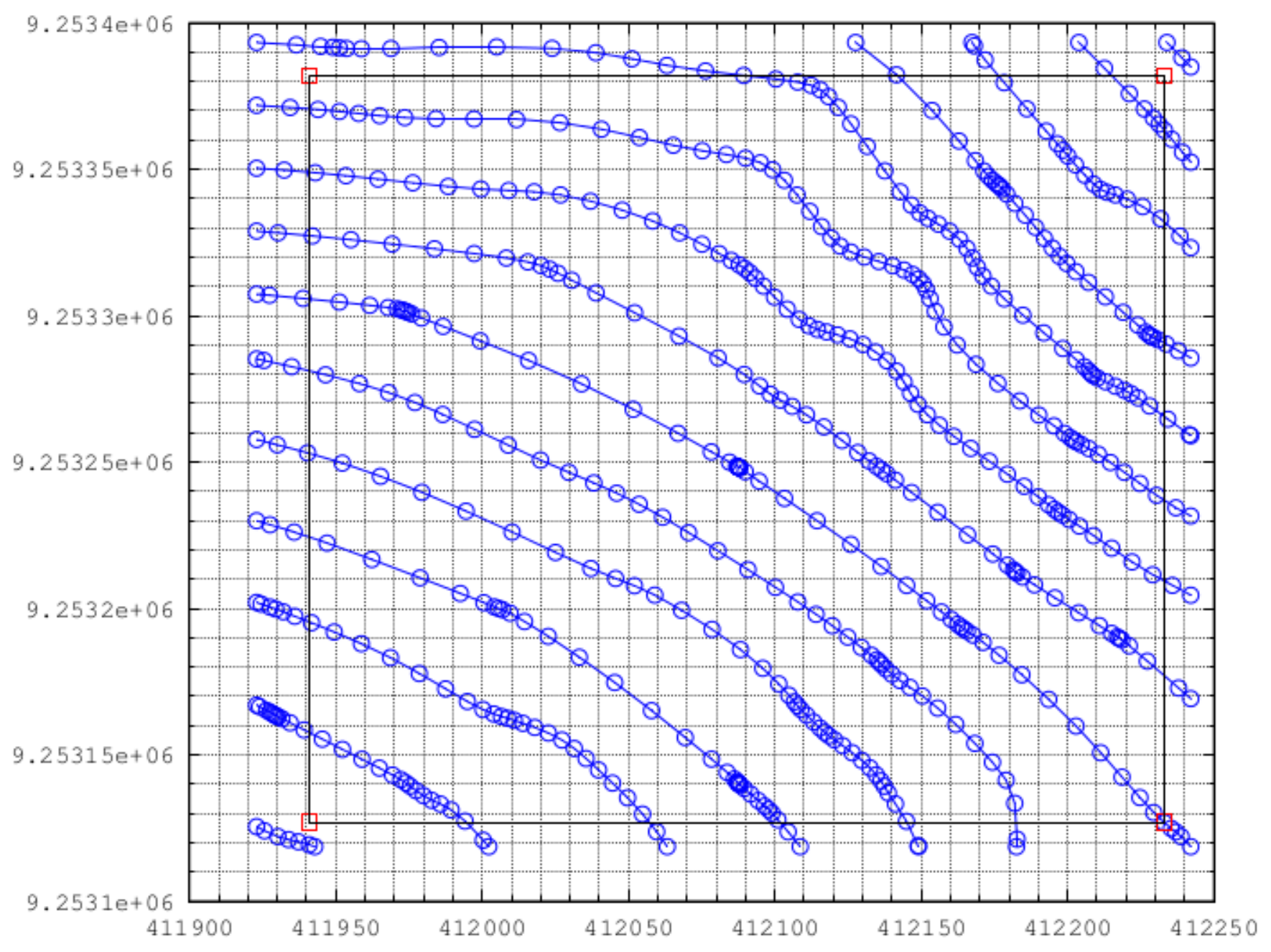}}
\caption{2D-rendering of a level curves-file.}
\label{fig:one}
\end{figure}

Anyway, the regularity of a plot extends to the terrain. This is because the UAV flies over a rectangular area that is just big enough to contain the plot. Therefore, the $(x,y)$-coordinates within any level curves-file fit in a rectangle. See our 2D-rendering of {\tt level\_curves1.txt} in Figure~\ref{fig:one}.

In Figure~\ref{fig:one} each tiny circle is centred at a certain $(x,y)$ of the data file. Connected circles have the same $z$-coordinate. This example has a very simple plot, which is the inner rectangle from $x=411,940.91$ to $x=412,232.99$ and from $y=9.25313e6$ to $y=9.25338e6$. Vertices of a plot are always marked with tiny squares.

As explained before, we just need to interpolate a sample of the $(x,y,z)$-points. In our programme the extreme coordinates of the terrain are called {\tt Xmax}, {\tt Xmin}, {\tt Ymax} and {\tt Ymin}. In practice, we get a 2D-polynomial out of a subgrid of at most $5\cdot5=25$ points equally spaced in $[Xmin,Xmax]\times[Ymin,Ymax]$. 

Hence the subgrid is an $N\times M$ matrix, and for each point $(x(i,j),y(i,j))$ of the subgrid there is a closest pair $(x,y)$ in the level curves-file, where $1\le i\le N\le 5$ and $1\le j\le M\le 5$. We take its corresponding $z$ to define $Z(i,j)$. Finally, we get the terms of our polynomial $z(x,y)$ as $q(i,j)x^{M-j}y^{N-i}$, where the coefficients $q(i,j)$ are obtained in Algorithm~\ref{alg:one}.

\begin{algorithm}
\SetAlgoNoLine
\KwIn{M,N,x,y,Z}
\KwOut{q}
\For{i=1:N}{p(i,:)=polyfit(x(i,:),Z(i,:),M-1)\;}
\For{j=1:M}{q(:,j)=polyfit(y(:,j),p(:,j),N-1)\;}
\caption{Obtaining the coefficients $q(i,j)$ of the polynomial $z(x,y)$}
\label{alg:one}
\end{algorithm}

Notice that we use the Matlab function {\tt polyfit} twice in order to get a single 2D-polynomial $z(x,y)$. Of course, {\tt polyfit} is devoted to 1D-interpolation. However, its double application is feasible because we have at most $25$ points.

\begin{figure}
\centerline{\includegraphics[scale=0.5]{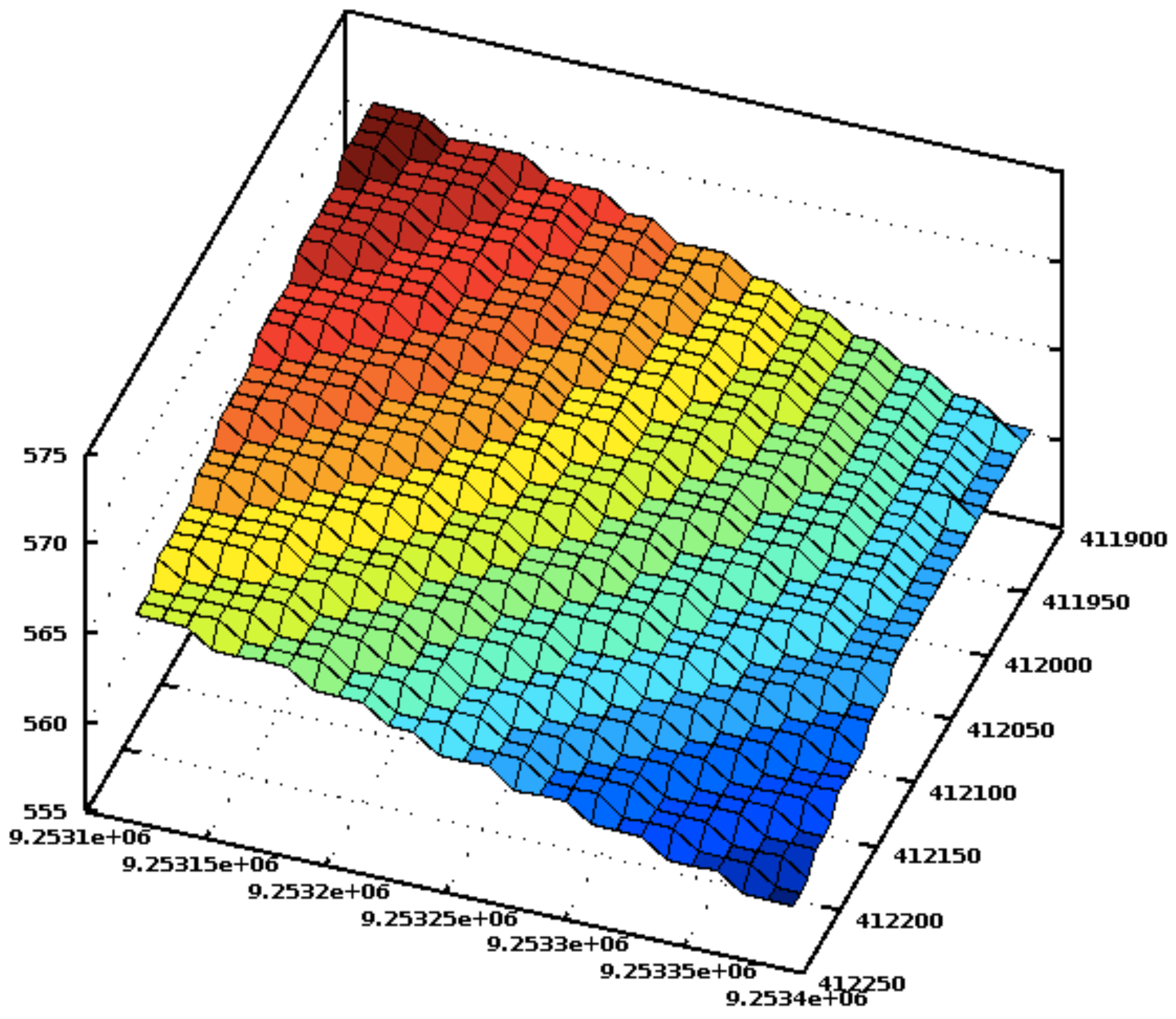}}
\caption{3D-rendering of a level curves-file.}
\label{fig:two}
\end{figure}

\begin{figure}
\centerline{\includegraphics[scale=0.5]{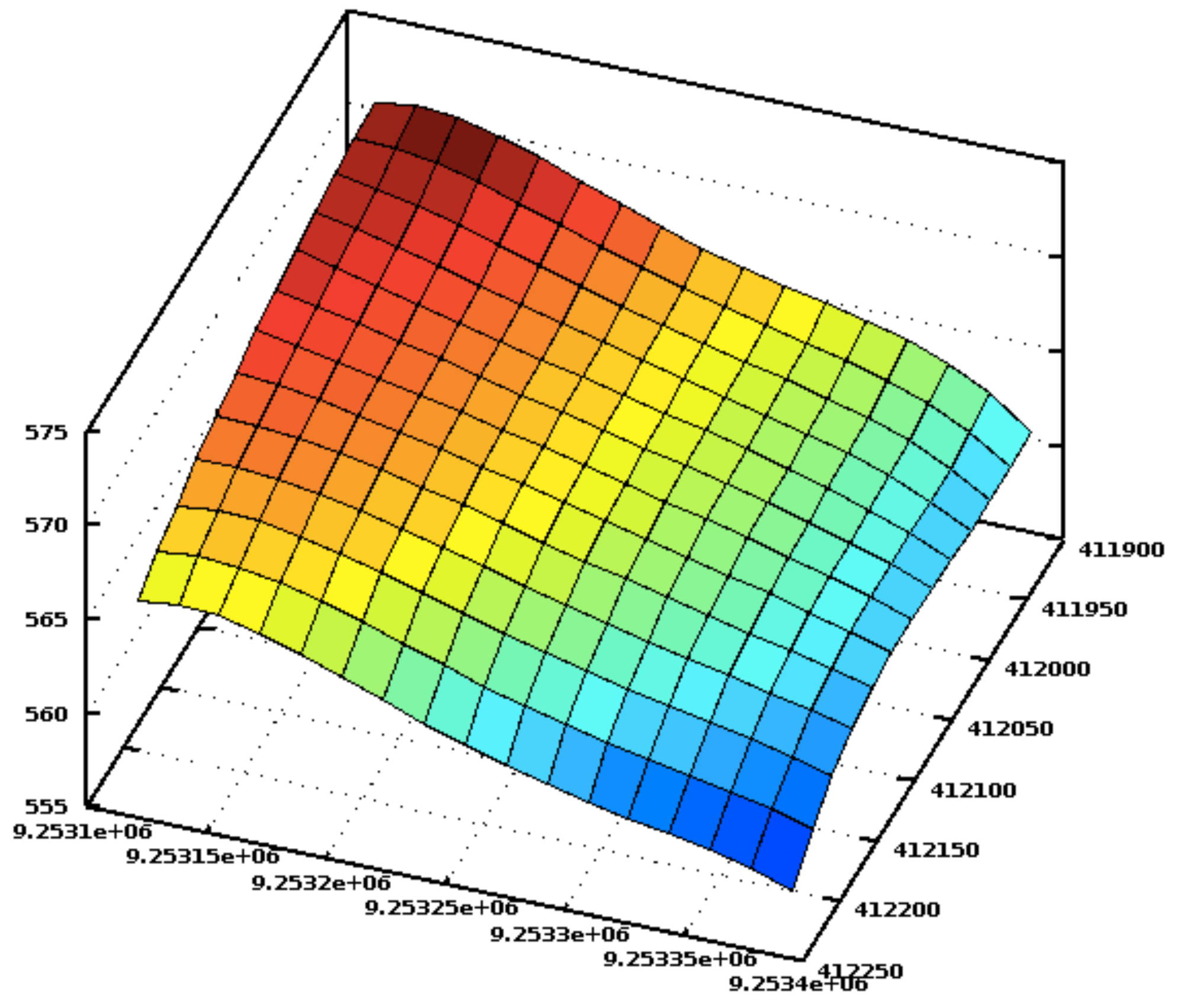}}
\caption{Plotting the polynomial $z(x,y)$ with the Matlab function {\tt surf(x,y,z)}.}
\label{fig:three}
\end{figure}

Figures~\ref{fig:two} and \ref{fig:three} were obtained from {\tt level\_curves1.txt}, and they exemplify how well the input data are reproduced by Algorithm~\ref{alg:one}.

\section{The Master Line}
\label{ml}

As remarked in the Introduction by condition (3), which is due both to agricultural and machine reasons, plantation lines should be parallel. Thus a single plantation line determines all the others. In the case of sugar cane, neighbouring plantation lines must be $3m$ distant from each other.

Because of condition (2), namely ``the tractor cannot make curves that are under $50m$ in radius'', we could start with a small curve that clings to the plot boundary. If it is an arc $C$ (of circumference) with radius $50m$, then all parallel lines will be wider.

Suppose that the plot boundary verifies a special condition explained in Subsection~\ref{eqterr}, namely ``both ways along the plot boundary from $H$ to $L$ are monotonically decreasing''. Thus $C$ should go round $H$, but even in this case there are infinitely many choices of an arc like that. 

This contrasts with condition (4): ``there must be the least possible number of plantation lines''. Hence our strategy is to trace the longest plantation line that passes through the middle of the plot under conditions (1)-(5). This will be called {\it the master line}. 

Due to condition (1), there must be little variation of the $z$-coordinate along the master line. In particular, condition (5) must hold. After having drawn our first choice, we can gradually perturb our master line in order to obtain the least number of parallels, as required by condition (4).

One detail that we have omitted so far: even on the most modern driverless tractors you still cannot programme a 3D-plantation line. This is just because the plot region must have a gentle slope. Moreover, condition (1) implies that the master line is almost planar. Therefore, if you programme a 2D-line on the tractor it will already follow a 3D-way by crossing the level curves.

Mathematically, the master line can be defined as a smooth regular curve $C:[0,\ell]\to\mathbb{R}^2$, where $\ell$ is its length. Without loss of generality, $C(t)=(x(t),y(t))$ is parametrized by its arclength. Its osculating circle has radius greater than $50$ for all $t\in[0,\ell]$. This implies that the parallels $C(t)+s(-y'(t),x'(t))$ are smooth and regular for $-50<s<50$. 

Now one must take enough parallels to cover the whole plot and guarantee that in its inside none of them will have an osculating radius under $50$. It is indeed a laborious task in the case of a general curve $C$. 

\begin{figure}[ht!]
\centerline{\includegraphics[scale=0.7]{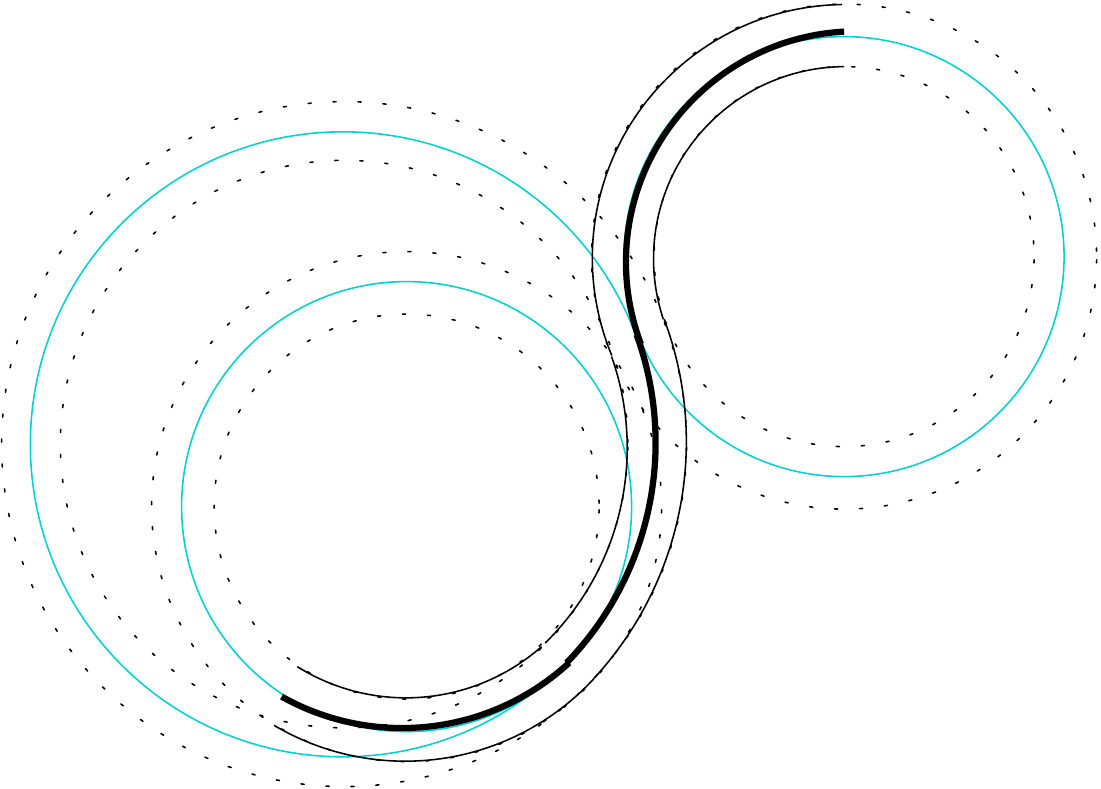}}
\caption{Parallel curves to an apw-master line.}
\label{fig:four}
\end{figure}

But suppose that $[0,\ell]$ admits a partition $0<\ell_1<\ell_2<\dots<\ell_k=\ell$ such that the restriction of $C$ to the sub-intervals is an arc of circumference. Two consecutive arcs must be tangent because $C$ is smooth. We call it {\it arc-piecewise} (apw). In this case, a classical result of Geometry states that all parallels to $C$ are also apw and their osculating centres will remain fixed. See Figure~\ref{fig:four}.

If a curve is apw, its osculating radius is always the same of the local arc. In our programme we obtain a master line that is apw. Since its parallels have all the same osculating centres, we first guarantee that none of them is in the plot. Afterwards we take the one with minimum distance $d$ to the plot. If $d<50$ we gradually widen the corresponding arc of $C$ until that centre is sufficiently moved away, namely $d\ge 50$. The process is repeated until all centres are farther than $50$ from the plot.  

This procedure is easy to implement when the master line consists of a single arc. We can just fix an internal point of the arc and change its extremities a bit towards alignment. Such a task will be called ``compute new extremities'' in Algorithm~\ref{alg:three} (see below). 

\section{Computing The Master Line for a Convex Plot}
\label{mlcp}

As described in Section~\ref{ml}, the master line is the longest plantation line that passes through the middle of the plot under conditions (1)-(5). Herewith we explain our algorithm for the case of a convex plot. Concave plots will be discussed in Section~\ref{mlcc}.

The vertices of the plot are stored in the variable {\tt pts}. They determine a polygon, which is the boundary of the plot. We need to go round the polygon and find the points that attain $H$ and $L$. If there are more than two, we take the pair with the farthest detachment. 

In order to inspect several points of the boundary, we go round it at a pace of at most $10m$ step. All these step-points are stored in the variable {\tt Pts}, which contains {\tt pts}.

The corresponding $z$-values on {\tt Pts} are computed via the 2D-polynomial $z(x,y)$ and stored in {\tt zc}. We have $H=\max(zc)$ and $L=\min(zc)$ attained at {\tt Pts(h)} and {\tt Pts(l)}, respectively. The ``middle of the plot'' is then {\tt mp}, as in Algorithm~\ref{alg:two}. The interpolating polynomial $z(x,y)$ at {\tt mp} gives {\tt zp}, and then we look for another two points on the boundary that are at level {\tt zp+de} and are the farthest from each other. We begin with {\tt de} $=2$ because in practice more than $2m$ already violates condition (1). 

\begin{algorithm}
\SetAlgoNoLine
\KwIn{H,L,Pts,z}
\KwOut{mp,zp}
zc=z(Pts)\;
[L,l]=min(zc)\;
[H,h]=max(zc)\;
mp=(Pts(h)+Pts(l))/2\;
zp=z(mp)\;
\caption{Obtaining the middle point of the plot}
\label{alg:two}
\end{algorithm}

\begin{figure}[ht!]
\centerline{\includegraphics[scale=0.45]{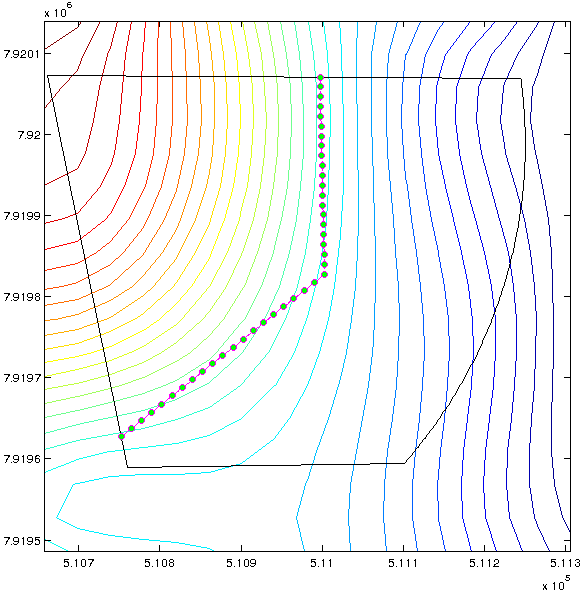}}
\caption{A prototype of the master line.}
\label{fig:five}
\end{figure}

Once we get these three points, a prototype of the master line is ready for us to begin with. Such a prototype is never shown to the user, but it is exemplified by {\tt plot3.txt} in Figure~\ref{fig:five}. 

The level curves in Figure~\ref{fig:five} are of the polynomial $z(x,y)$, not of the level curves-file. Our programme always shows both in separate windows for comparison.

The initial choice {\tt de} $=2$ also guarantees that the three strategic points of the prototype will never be collinear ({\tt mp} and the extremities). Hence, they determine an arc of circumference. This arc is then adjusted to optimize the master line according to conditions (1)-(5) as described in Algorithm~\ref{alg:three}. Of course, we use condition (3) in order to count the number of plantation lines that cover the plot. This number is stored in the variables {\tt nbf} and {\tt naf}, which count it before and after the while-loop, respectively.

\begin{algorithm}
\SetAlgoNoLine
\KwIn{de,master line,Pts}
\KwOut{optimized master line}
$nbf=999$; $naf=999$\;
$c = $ centre of master line\;
$d=\min|c-Pts|$\;
\While{( ($d<50$ or $naf\le nbf$) and $de>0$ )}{
$nbf=naf$\;
$de=de/2$\;
compute new extremities\;
$c = $ centre of new master line\;
$d=\min|c-Pts|$\;
$naf=( \max|c-Pts| - d )/3$\;}
\eIf{$\max|angle|\le5^\circ$}{master line = optimized master line\;}
{printf(``Angle Not OK'')\;}
\caption{Optimizing the master line}
\label{alg:three}
\end{algorithm}

In the next section we shall apply Algorithm~\ref{alg:three} again. Profiting from already implemented codes is another reason why our programme was written with only 573 lines.

\section{Computing The Master Line for a Concave Plot}
\label{mlcc}

As explained in Section~\ref{mlcp}, the plot boundary is a polygon. Its vertices are stored in the variable {\tt pts}. From classical Geometry we know that a set is convex exactly when it coincides with its convex hull. Our sub-programme {\tt cvxhull} computes it for {\tt pts}. If the sets do not coincide, then the plot is concave.

\begin{figure}
\centerline{\includegraphics[scale=0.45]{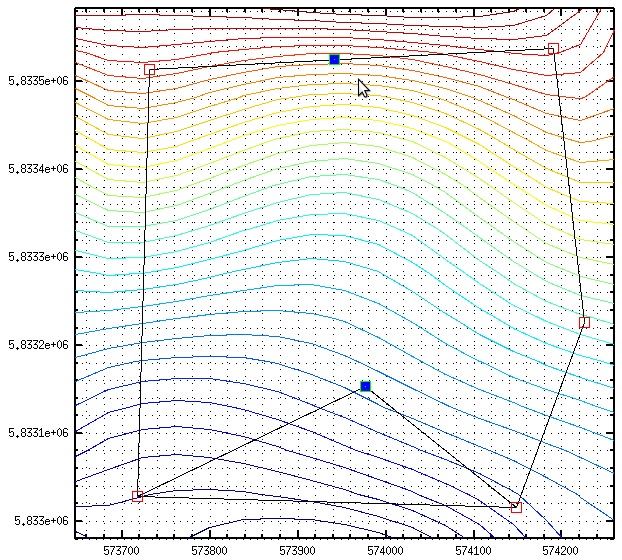}}
\caption{Choosing strategic points to subdivide the plot.}
\label{fig:six}
\end{figure}

\begin{figure}
\centerline{\includegraphics[scale=0.7]{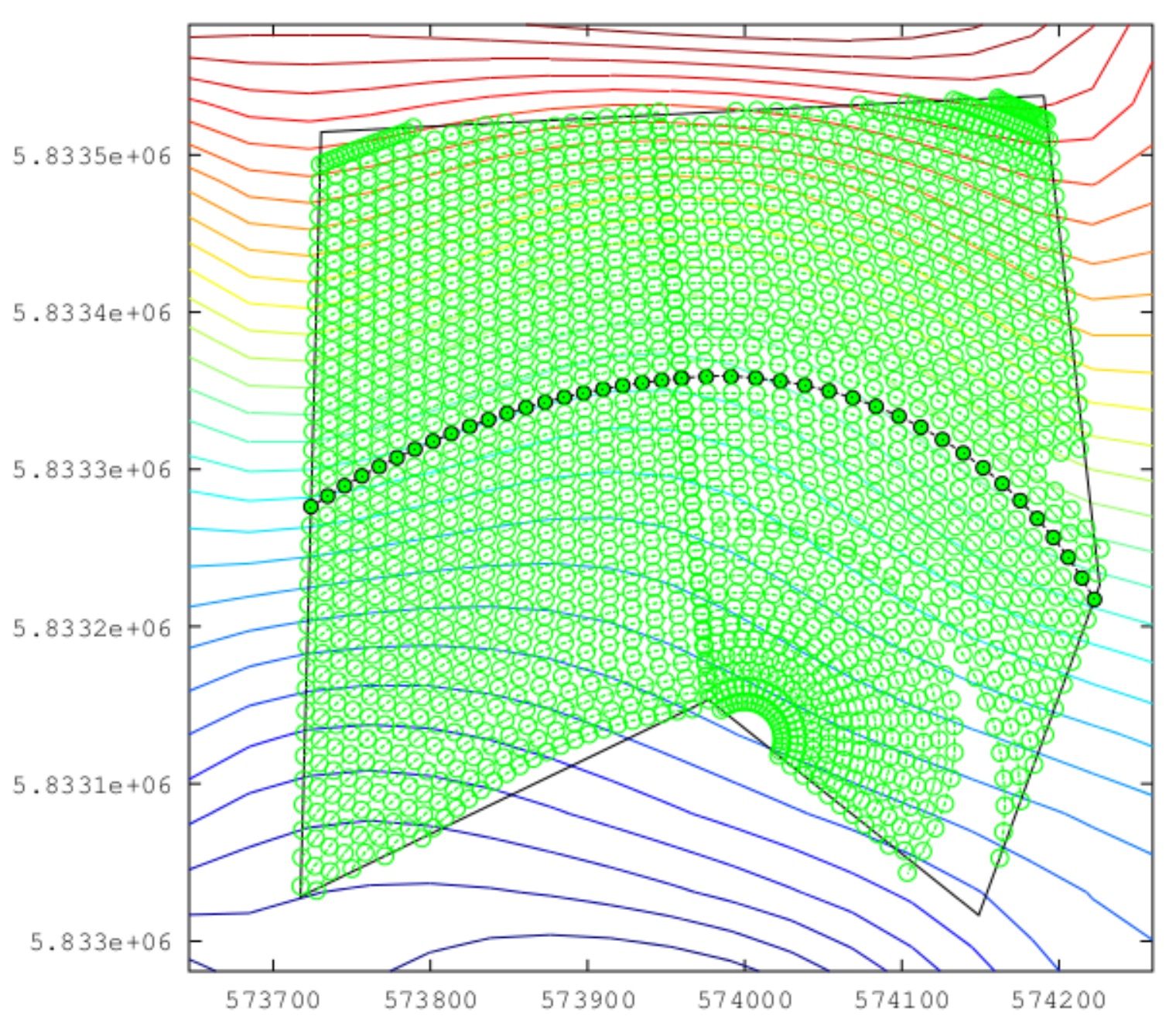}}
\caption{A bad choice of extra points from Figure~\ref{fig:six}.}
\label{fig:seven}
\end{figure}

It is worthwhile commenting that {\tt cvxhull.m} has only 30 lines of source code. As a matter of fact, this is mostly due to the plots being handled in 2D. Generating the convex hull of a spatial set is far much harder a task (see \cite{Gao}).

Concave plots can be subdivided into two or more convex regions. It is precisely at this moment that our programme asks the user to intervene. They have to choose strategic points that subdivide the plot into ``nice'' convex regions. Intuitively, ``nice'' means that we should get an apw-master line that verifies conditions (1)-(5) for the plot as a whole.

Namely, there are infinitely many ways of subdividing the plot into convex regions. However, instead of considering them separately, they must be viewed as parts of an entire plot. Since this is highly intuitive we do believe that unsupervised methods will fail to make good subdivisions.

In practice the user will have to select at most three pairs of points. It is done by clicking the the mouse on chosen points of the polygon. This task becomes trivial with time, but the user can save the chosen points as soon as a good subdivision is found. Figure~\ref{fig:six} exemplifies this task on {\tt level\_curves2.txt} with a very simple concave plot. 

The choice from Figure~\ref{fig:six} resulted in a bad master line, as depicted in Figure~\ref{fig:seven}. 

That is why we have not saved the chosen points. After some tries the best choice was finally found and stored in {\tt eplot2.txt} (contained in our zip-file). 

\section{Conclusions}

We hope to have contributed with the semi-automation of a task that has been performed manually for many decades. Of course, our programme has limitations and {\tt level\_curves4.txt} shows an example. 

\begin{figure}[ht!]
\centerline{\includegraphics[scale=0.7]{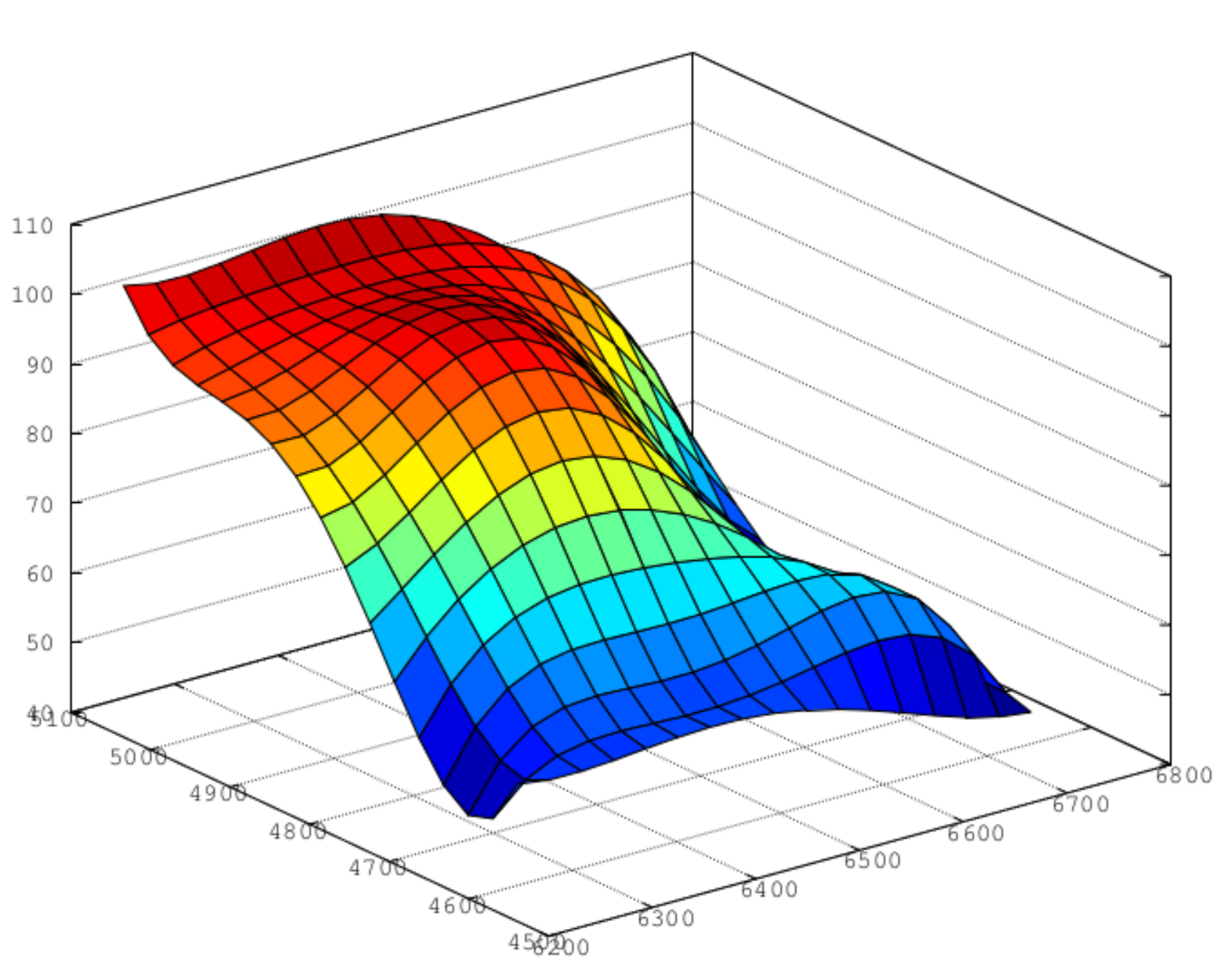}}
\caption{A slight hill inside the plot.}
\label{fig:eight}
\end{figure}

However, Figure~\ref{fig:eight} indicates a problem that commonly remains unsolved even if tackled manually by good experts: the inside of the plot is hilly. In practice, such a plot is subdivided in two subplots by a lane and the tractor has to perform man\oe uvres on the lane when it passes from one region to the other.

It is still unclear whether the ideal lane can be found by a computer programme. As a matter of fact, we believe that some tasks will ever remain to be accomplished manually. This is not exclusive to Agriculture. For instance, Computer Aided Diagnosis (CAD) has been an indispensable help in Medicine for many decades already. In the case of digital(ized) mammographies, CAD systems can detect and classify nodules as normal, benign or malignant, this latter being the only kind that requires extraction. The automatic classification makes uses of several techniques. One of them is the Wavelet Transform (see \cite{Taswell}). 

Anyway, no matter how good CAD systems are, they in fact serve as tool to supplementary analysis that helps radiologists achieve preciser diagnoses. The ultimate conclusions must still rely on these professionals. For details, see \cite{Zanka1,Zanka3,Zanka4}.  

\bibliographystyle{plain}
\bibliography{toms}
\end{document}